\documentclass[10pt,conference]{IEEEtran}

\usepackage{url}
\usepackage{amsmath,amssymb,amsfonts}
\usepackage[linesnumbered,ruled]{algorithm2e}
\usepackage{algorithmic}
\usepackage{graphicx}
\usepackage{textcomp}
\usepackage{xcolor}
\usepackage{tipa}
\usepackage{tabularx}
\usepackage{color}
\usepackage{soul}
\usepackage{tcolorbox}
\usepackage{caption}
\usepackage{subcaption}
\usepackage{mwe}
\usepackage{enumitem}
\usepackage[numbers,sort&compress]{natbib}
\usepackage{multirow}

\newcommand{\resultBox}[1]{\begin{tcolorbox}\textbf{#1}\end{tcolorbox}}
\newcommand{\ea}{\textit{et al.}}

\newcommand{\rqi}{{Which goals of defect prediction models that practitioners considered the most useful?}}

\newcommand{\rqii}{Which model-agnostic techniques are the most preferred by practitioners to understand defect prediction models and their predictions?}

\begin{document}

\title{Practitioners' Perceptions of the Goals and Visual Explanations of Defect Prediction Models}


\author{%
    \IEEEauthorblockN{Jirayus Jiarpakdee, Chakkrit (Kla) Tantithamthavorn,  John Grundy}
    \IEEEauthorblockA{Monash University, Melbourne, Australia.
    } 
}

\maketitle




\begin{abstract}

Software defect prediction models are classifiers that are constructed from historical software data. 
Such software defect prediction models have been proposed to help developers optimize the limited Software Quality Assurance (SQA) resources and help managers develop SQA plans.
Prior studies have different goals for their defect prediction models and use different techniques for generating visual explanations of their models.
Yet, it is unclear what are the practitioners' perceptions of (1) these defect prediction model goals, and (2) the model-agnostic techniques used to visualize these models.
We conducted a qualitative survey to investigate practitioners' perceptions of the goals of defect prediction models and the model-agnostic techniques used to generate visual explanations of defect prediction models.
We found that 
(1) 82\%-84\% of the respondents perceived that the three goals of defect prediction models are useful;
(2) LIME is the most preferred technique for understanding the most important characteristics that contributed to a prediction of a file, while ANOVA/VarImp is the second most preferred technique for understanding the characteristics that are associated with software defects in the past.
Our findings highlight the significance of investigating how to improve the understanding of defect prediction models and their predictions. 
Hence, model-agnostic techniques from explainable AI domain may help practitioners to understand defect prediction models and their predictions.

\end{abstract}


\section{Introduction} 

Software defects are prevalent, but hard to predict~\cite{menzies2007data} and to prevent~\cite{Nagappan2005, nagappan2006mining}.
Thus, prior studies developed defect prediction models from historical software data using a statistical or machine learning model for various purposes (e.g., prediction and explanations), which addresses various goals.
First is to \emph{predict} the likelihood of a file~\cite{menzies2007data,hall2012systematic} or a commit~\cite{Kamei2013,yang2016effort,pornprasit2021jitline} being defective in the future.
Second is to \emph{understand} the characteristics that are associated with software defects in the past~\cite{Nagappan2005,nagappan2006mining}.
Third is to \emph{explain} their prediction about why a particular file is predicted as defective~\cite{jiarpakdee2020modelagnostic,peng2020defect,peng2020improve}. 

Prior studies hypothesized that the predictions could help practitioners prioritize the limited inspection effort on the most risky files~\cite{zimmermann2007predicting,menzies2007data,hata2012bug,wattanakriengkrai2020linedp,yang2016effort}, while the insights derived from defect prediction models could help managers chart appropriate quality improvement plans~\cite{shrikanth2020assessing,Nagappan2005,nagappan2006mining}.
Recently, Wan~\ea~\cite{wan2018perceptions} conducted a survey study with practitioners to investigate their perceptions of defect prediction models.
However, Wan~\ea~\cite{wan2018perceptions} only focused on the prediction goal, while the other two goals (i.e., understanding models and explaining the predictions) have not been investigated.
A better understanding of the practitioners' perceptions will help the research community to better understand practitioners' needs, allowing researchers to orient appropriate efforts for the design and the development of the next-generation of defect prediction models.

Prior studies used various model-agnostic techniques from Explainable AI~\cite{guidotti2018survey,tantithamthavorn2020explainable, wattanakriengkrai2020linedp,pornprasit2021jitline,peng2020improve,jiarpakdee2020modelagnostic}---i.e., techniques for generating explanations of defect prediction models and their predictions to help practitioners understand defect prediction models and their predictions~\cite{Nagappan2005,Bird2011a,thongtanunam2016revisiting, thongtanunam2017review,jiarpakdee2020modelagnostic}.
In fact, different model-agnostic techniques generate different key information, e.g., importance scores and relationship. 
However, none of the prior studies investigates which model-agnostic techniques are considered as the most preferred by practitioners to generate visual explanations.
A better understanding of practitioners' perceptions of the visual explanations of defect prediction models is needed to guide researchers to devise novel visualization techniques that suit practitioners' needs.

In this paper, we conducted a qualitative survey to understand the practitioners' perceptions of the goals of defect prediction models and the model-agnostic techniques for generating visual explanations of defect prediction models.
The analysis of related work led us to focus on three goals of defect prediction models: (1) prioritizing the limited SQA resources on the most risky files; (2) understanding the characteristics that are associated with software defects in the past; and (3) understanding the most important characteristics that contributed to a prediction of a file. 
We asked respondents to assess the perceived usefulness and their willingness to adopt defect prediction models.
Then, we asked the respondents to describe the positive and negative impacts if these defect prediction models were adopted.

Guided by the analysis of related work, we focused on 8 model-agnostic techniques used for generating visual explanations -- ANOVA, Variable Importance, Partial Dependence Plots, Decision Tree, LIME~\cite{ribeiro2016should}, BreakDown~\cite{gosiewska2019ibreakdown}, SHAP~\cite{lundberg2017unified}, and Anchor~\cite{ribeiro2018anchors}.
We asked our survey respondents to assess each visual explanation generated by these techniques along three dimensions, i.e., information usefulness, information insightfulness, and information quality.
We then asked the respondents to describe the positive and negative feedback of each visual explanation of defect prediction models.
Through a qualitative analysis of open-ended and closed-ended questions of 50 software practitioners we found that:

\begin{itemize}
    \item \textbf{82\%-84\% of the respondents perceived that the three goals of defect prediction models are useful and 74\%-78\% of them are willing to adopt them}. This was especially true for respondents who use Java, have little years of experience, and work in a large team size (more than 100 people).
    This finding highlights that not only defect predictions but also the other two goals (i.e., understanding defect prediction models and their predictions) receive similar perceptions of usefulness and willingness to adopt with no statistically significant difference.
    \item \textbf{LIME is the most preferred technique for understanding the most important characteristics that contributed to a prediction of a file with an agreement percentage of 66\%-78\% along three dimensions.} 
    ANOVA/VarImp is the second most preferred technique for understanding the characteristics that are associated with software defects in the past with an agreement percentage of 58\%-70\% along three dimensions.
\end{itemize}

Based on these findings, future research (1) should put more effort on investigating how to improve the understanding of defect prediction models and their predictions; and (2) can use ANOVA/VarImp and LIME model-agnostic techniques from explainable AI domain to understand defect prediction models and their predictions.
We also discuss key lessons learned and open questions for developing the next-generation of defect prediction models, e.g., how to develop the highly-scalable human-in-the-loop defect prediction models at the lowest implementation cost, while maintaining its explainability.

The main contributions of this paper are as follows:

\begin{itemize}
\item We conducted a qualitative survey on practitioners' perceptions of the goals of defect prediction models and the model-agnostic techniques for generating visual explanations of defect prediction models. We also provided a detailed replication package and a tutorial in Zenodo~\cite{onlineappendix}.
\item We investigated the key factors that impact practitioners' perceptions of the goals of defect prediction models and the model-agnostic techniques for generating visual explanations of defect prediction models.
\item We identified a key set of implications for researchers including  open questions for future research on designing and developing the next-generation defect models.
\end{itemize}

\section{Related Work \& Research Questions}\label{sec:background}

We first summarize key related work to identify (i) the key goals of developing defect prediction models, and (ii) the model-agnostic techniques that are used to generate visual explanations.
We then motivate our research questions based on the analysis of related work.

\subsection{Related Work}

{We collected the titles of full research track publications that were published in the top SE venues (i.e., TSE, ICSE, EMSE, FSE, and MSR) during 2015-2020 from IEEE Xplore, Springer, and ACM Digital Library (as of 11 January 2021). 
These venues are premier publication venues in the software engineering research community. 
We used the ``defect", ``fault", ``bug", ``predict", and ``quality" keywords to search for papers about defect prediction models. 
This led us to a collection of 2,890 studies.
Since studies may use several keywords that match with our search queries and appear consistently across the search results, we first identified and excluded duplicate studies.
We found that 1,485 studies are duplicated and thus are excluded (1,405 unique studies).
Then, we manually read the titles and abstracts of these papers to identify whether they are related to defect prediction models. 
For each paper, we manually downloaded {each paper as a pdf file} from IEEE Xplore, Springer, and ACM Digital Library.
We identified 131 studies that are relevant to software defects prediction.
Of the 131 studies, we excluded 7 studies that are not primary studies of defect prediction models} (e.g., secondary studies \cite{zou2018practitioners,hosseini2017systematic}).
Then, we excluded 28 studies that are not full-paper and peer-reviewed defect prediction studies (i.e., short, journal first, extended abstract papers).
Finally, we selected a total of 96 primary full-paper and peer-reviewed defect prediction studies.



We read each of them to identify their goals of developing defect prediction models and identify the model-agnostic techniques that are used to generate visual explanations.
We then further group the goals of developing defect prediction models using the Open Card Sorting approach. 
First, we list all the goals of developing defect prediction models and categorize these goals based on how such models are used in each study.
Then, we discuss the inconsistency among the authors to reach the final set of categories.
Based on the selected 96 studies, we identify the 3 goals of developing defect prediction models and 8 model-agnostic techniques used to generate visual explanations.
Guided by the Guidotti~\ea's Taxonomy~\cite{guidotti2018survey}, we classify each technique according to types of model-agnostic techniques (i.e., interpretable models vs. post-hoc explanations) and the granularity levels of explanations (i.e., model explanation and outcome explanation~\cite{jiarpakdee2020modelagnostic}). 

%



\renewcommand\arraystretch{1.1}

\newcolumntype{T}{>{\centering\arraybackslash}m{0.09\textwidth}}
\newcolumntype{I}{>{\centering\arraybackslash}m{0.24\textwidth}}
\newcolumntype{R}{>{\centering\arraybackslash}m{0.21\textwidth}}

\begin{table*}[t]
\centering
\caption{A summary of type, granularity, and key information of the model-agnostics techniques that are used in defect prediction studies~\cite{guidotti2018survey, jiarpakdee2020modelagnostic}, and an example of positive and negative feedback from practitioners (RQ2).}

\resizebox{\textwidth}{!}{
\begin{tabular}{l|T|p{2cm}|p{3.1cm}||R|R}
\hline
\textbf{Goal}                     & \multicolumn{1}{c|}{\textbf{Technique}} & \multicolumn{1}{c|}{\textbf{Types/Granularity}} & \multicolumn{1}{c||}{\textbf{Information}}                  & \multicolumn{1}{c|}{\textbf{Positive feedback}}                                                      & \multicolumn{1}{c}{\textbf{Negative feedback}} \\ \hline 
\multirow{3}{*}{\textbf{\begin{tabular}[c]{@{}l@{}}(Goal 2)\\Understanding the\\characteristics that\\are  associated with\\software defects\\in the past\end{tabular}}} & \textbf{Anova/ VarImp}  & A Post-hoc Explainer for Model Explanation                 & (1) The importance scores of each feature                  & R50 (\emph{``Explains risk values of several factors.''})                                                & R34 (\emph{``It is a very basic plot which could even have been a table.''}) \\ \cline{2-6} 
 & \textbf{Partial Dependence Plot (PDP)} & A Post-hoc Explainer for Model Explanation  & (1) The relationship of each feature on the outcome & 
 R5 (\emph{``I like that I can see a trending pattern to read with ease.''}) & 
 R27 (\emph{``Easy to visualize but hard to process the info.''}) \\ \cline{2-6} 
 & \textbf{Decision Tree} & An Interpretable Model for Model Explanation                 & (1) The decision rule of each feature                      &
R47 (\emph{``Very analytical, strong with engineers, flow chart.''}) & 
R34 (\emph{``Very difficult to read especially when there are multiple parameters and the tree gets un-manageable.''}) \\ \hline
\multirow{19}{*}{\textbf{\begin{tabular}[c]{@{}l@{}}(Goal 3)\\Understanding the\\most important\\characteristics that\\contributed to a\\prediction of a file\end{tabular}}} & \multirow{9}{*}{\textbf{LIME}} & {\multirow{9}{*}{\begin{tabular}[l]{@{}l@{}}A Post-hoc\\Explainer for\\ Outcome \\ Explanation\end{tabular}}}         & (1) The prediction of the local models                     & \multirow{8}{*}{
    \textit{\begin{tabular}[c]{@{}l@{}}
        R1 (``Risk scores are visually \\ oriented and users will \\ understand  it faster.'')\\ R50 (``Easy to understand, we \\ just want to know what's \\ already ok and what's need to \\ improve.'')
    \end{tabular}}
} 
& \multirow{8}{*}{
    \textit{\begin{tabular}[c]{@{}l@{}}
        R9 (``... But it also could be time\\consuming to review depending \\ on if there are many different \\ files to review.'')\\ R50 (``... too large and too tired \\ to read 10 charts for 10 files.'')
    \end{tabular}}
}                            \\ \cline{4-4}
 & & &  (2) The decision point of each feature & & \\ \cline{4-4}
 & & & (3) The supporting scores of each feature & & \\ \cline{4-4}
 & & & (4) The contradicting scores of each feature & & \\ \cline{4-4}
 & & & (5) The actual feature value of that instance & & \\ \cline{2-6} 
 & \multirow{6}{*}{\textbf{SHAP}} & \multirow{6}{*}{\begin{tabular}[l]{@{}l@{}}A Post-hoc\\Explainer for\\Outcome \\ Explanation\end{tabular}} & (1) The prediction of the local models & 
 \multirow{10}{*}{
 \textit{\begin{tabular}[c]{@{}l@{}}
        R5 (``... It does highlight in the\\explanation of what everything \\ is, which is nice. It gives \\ insightful data on what to look \\ out for.'')\\ R14 (``It describes the \\ involvement of each attribute \\ more clearly.'')
    \end{tabular}}
 }                               & \multirow{10}{*}{
 \textit{\begin{tabular}[c]{@{}l@{}}
        R9 (``This has a lot of information\\which can be quite useful but lacks\\a clean readability. It can be quite\\time-consuming to read this graph.'')\\ R27 (``Takes a bit to figure out \\ what is going on.'')
    \end{tabular}}
 } \\ \cline{4-4}
 & & &  (2) \% of contribution for each feature to the final probability & & \\ \cline{4-4}
 & & & (3) The actual feature value of that instance & & \\ \cline{2-4}
 & \multirow{6}{*}{\textbf{BreakDown}}& \multirow{6}{*}{\begin{tabular}[l]{@{}l@{}}A Post-hoc\\Explainer for\\Outcome \\ Explanation\end{tabular}} & (1) The prediction of the local models & & \\ \cline{4-4}
 & & &  (2) \% of contribution for each feature to the final probability & & \\ \cline{4-4}
 & & & (3) The actual feature value of that instance & & \\ \cline{2-6} 
 & \textbf{Anchor}& \begin{tabular}[l]{@{}l@{}}A Post-hoc\\Explainer for\\Outcome \\ Explanation\end{tabular} & (1) The decision rule of each feature & R50 (\emph{``Exactly what I need, just small and short information on what to improve.''}) & R36 (\emph{``... doesn't provide a visual aid to put the numbers into perspective.''}) \\ \hline
\end{tabular}
}
\label{review-table}
\end{table*}



\subsubsection*{\textbf{Goal 1---Prioritizing the limited SQA resources on the most risky files}}

Software defects are prevalent in many large-scale software systems (e.g., 47K+ for Eclipse, and 168K+ for Mozilla)~\cite{lamkanfi2013eclipse}.
Developers have to exhaustively review and test each file in order to identify and localize software defects. 
However, given thousands of files in a software system, exhaustively performing SQA activities are likely infeasible due to limited SQA resources (e.g., in a rapid release setting).
Thus, prior studies use defect prediction models to predict the likelihood of a file being defective in the future.
Such predictions can be used to produce a ranking of the most risky files that require SQA activities.
Prior studies leveraged several Machine Learning approaches to develop defect prediction models to predict which files~\cite{zimmermann2007predicting,menzies2007data}, methods~\cite{hata2012bug}, lines~\cite{wattanakriengkrai2020linedp} are likely to be defective in the future.
For example, regression models~\cite{zimmermann2007predicting, Bird2011a,thongtanunam2016revisiting, thongtanunam2017review,Nagappan2005}, random forests~\cite{tantithamthavorn2015icse}, and deep learning~\cite{wang2016automatically}.

\subsubsection*{\textbf{Goal 2---Understanding the characteristics that are associated with software defects in the past}}
Numerous characteristics are associated with software quality.
For example, the static and dynamic characteristics of source code~\cite{hall2012systematic,zimmermann2007predicting}, software development practices (e.g., the amount of added lines)~\cite{misirli2016studying,Kamei2013}, organizational structures (e.g., social networks)~\cite{Nagappan2005}, and human factors (e.g., the number of software developers, code ownership)~\cite{Bird2011a,thongtanunam2016revisiting, thongtanunam2017review}.
Yet, different systems often have different quality-impacting characteristics.
Thus, prior studies used defect prediction models to better understand such characteristics that are associated with software defects in the past.
This understanding could help managers chart appropriate quality improvement plans.
Below, we summarize the four model-agnostic techniques that are used in prior studies to understand the characteristics that are associated with software defects in the past.

\emph{Analysis of Variance (ANOVA)} is a model-agnostic technique for regression analysis to generate the importance scores of factors that are associated with software defects.
ANOVA measures the importance of features by calculating the improvement of the Residual Sum of Squares (RSS) made when sequentially adding a feature to the model. 
\emph{Variable Importance (VarImp)} is a model-agnostic technique for random forests classification techniques to generate the importance scores of factors that are associated with software defects.
The VarImp technique measures the importance of features by measuring the errors made when the values of such features are randomly permuted (i.e., permutation importance).
Random forests also provides other variants of importance score calculations (e.g., Gini importance). 
In this paper, we choose the permutation importance technique to generate an example of VarImp visual explanation since we find that permutation importance is more robust to the collinearity issues~\cite{jiarpakdee2018impact}.
We note that the ANOVA and VarImp plots only indicate the importance of each feature, not the directions of the relationship of each feature i.e., positive or negative.

\emph{Partial Dependence Plot (PDP)}~\cite{friedman2001greedy} is a model-agnostic technique to generate model explanations for any classification models.
Unlike visual explanations generated by ANOVA and VarImp that show only the importance of all features, visual explanations generated by PDP illustrate the marginal effect that one or two features have on the predicted outcome of a classification model.

\emph{Decision Tree} or Decision Rule is a technique to generate tree-based model explanations. 
A decision tree is constructed in a top-down direction from a root node.
Then, a decision tree partitions the data into subsets of similar instances (homogeneous). 
Typically, an entropy or an information gain score are used to calculate the homogeneity of among instances.
Finally, the constructed decision tree can be converted into a set of if-then-else decision rules.

\subsubsection*{\textbf{Goal 3---Understanding the most important characteristics that contributed to a prediction of a file}}
Recently, Jiarpakdee~\ea~\cite{jiarpakdee2020modelagnostic} argued that a lack of explainability of defect prediction models could hinder the adoption of defect prediction models in practice (i.e., developers do not understand why a file is predicted as defective).
To address this challenge, Jiarpakdee~\ea~proposed to use model-agnostic techniques to generate explanations of the predictions of defect prediction models (i.e., what are the most important characteristics that contributed to a prediction of a file?).
Below, we summarize the four state-of-the-art model-agnostic techniques that were used in prior studies to understand the most important characteristics that contributed to a prediction of a file (i.e., LIME, BreakDown, SHAP, and Anchor).

\emph{Local Interpretability Model-agnostic Explanations (LIME)}~\cite{ribeiro2016should} is a model-agnostic technique to generate the importance score of the decision rule of each factor for any classification models.
The decision rule of each factor is discretized based on a decision tree.
LIME aims to generate supporting and contradicting scores which indicate the positive and negative importance of each feature for an instance.
For example, a LIME explanation for the LOC feature with an importance score of 40\% and a decision rule $\mathrm{LOC} > 100 => \mathrm{BUG}$ indicates that the condition of the file size that is larger than 100 LOCs would have 40\% contribution to the prediction that a file is defective.


\emph{BreakDown}~\cite{gosiewska2019ibreakdown} is a model-agnostic technique for generating probability-based explanations for each model prediction~\cite{staniak2018explanations}.
BreakDown uses the greedy strategy to sequentially measure the contributions of each feature towards the expected predictions.
For example, a BreakDown explanation for the LOC feature with an importance score of 40\% indicates that the actual feature value of 200 LOCs of the file would have 40\% contributions to the final prediction of this particular file as being defective.


\emph{SHapley Additive exPlanations (SHAP)}~\cite{lundberg2017unified} is a model-agnostic technique for generating probability-based explanations for each model prediction based on a game theory approach.
SHAP uses game theory to calculate the Shapley values (contributions) of each feature based on the decision-making process of prediction models.


\emph{Anchor}~\cite{ribeiro2018anchors} is an extension of LIME~\cite{ribeiro2016should} that uses decision rules to generate rule-based explanations for each model prediction.
The key idea of Anchor is to select if-then rules -- so-called anchors -- that have high confidence, in a way that features that are not included in the rules do not affect the prediction outcome if their feature values are changed. In particular,
Anchor selects only rules with a minimum confidence of 95\%, and then selects the rule with the highest coverage if multiple rules have the same confidence value.

\subsection{Research Questions}




\begin{figure}[t]
  \centering
  \includegraphics[width=.9\linewidth, trim={0 0 0 0}]{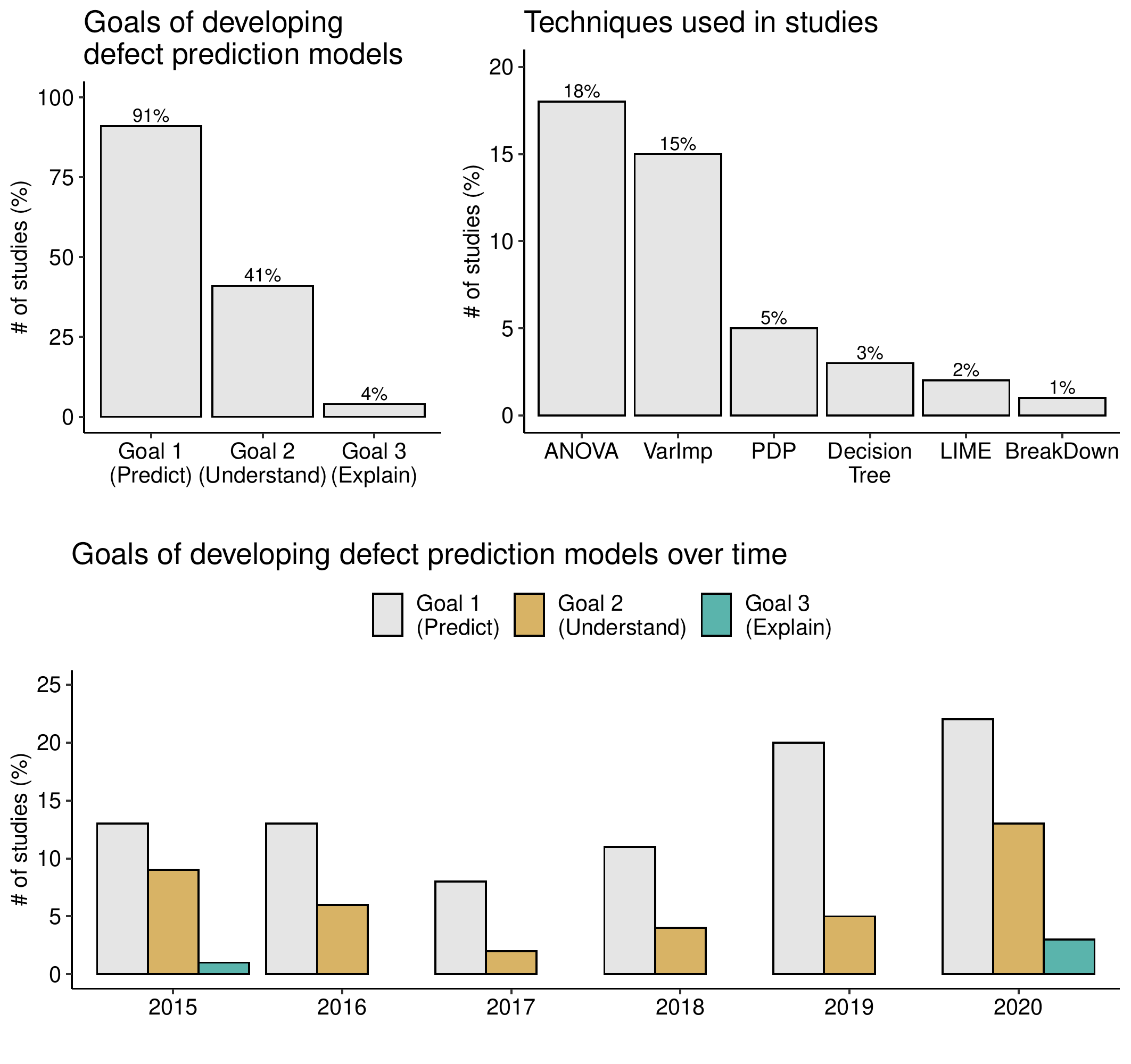}
\caption{The proportion of the goals of developing defect prediction models and the proportion of the model-agnostic techniques used in prior studies. We note that the summation of these percentage values does not add up to 100\% since a study may have multiple goals and may use multiple model-agnostic techniques.}
\label{fig:review-figures}
\end{figure}

As shown in Figure~\ref{fig:review-figures}, we found that most recent defect prediction studies focus on prioritizing the limited SQA resources. 
This led us to hypothesize that the prediction goal is perceived as more useful than the other two goals, i.e., understanding defect prediction models and their predictions.
However, it remains unclear how do practitioners perceive the three goals of defect prediction models.
Thus, we formulate the following research question:

\begin{figure}[t]
  \centering
  \includegraphics[width=\columnwidth]{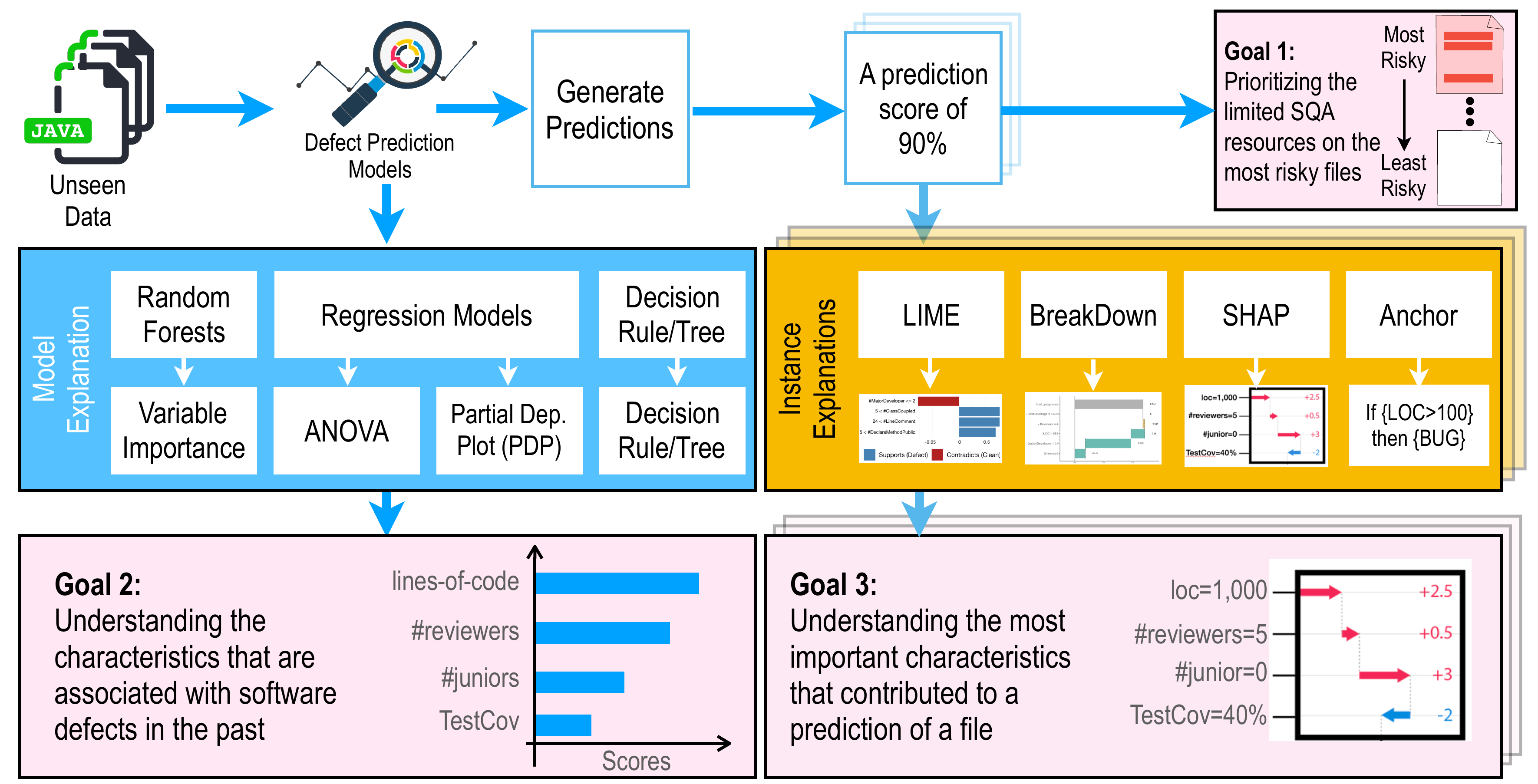}
  \caption{An illustrative overview of the goals of defect prediction models and the model-agnostic techniques for generating visual explanations of defect prediction models.}
  \label{framework}
\end{figure}


\resultBox{(RQ1) \rqi}

According to our analysis of related work, prior defect prediction studies also used model-agnostic techniques to generate visual explanations to help practitioners understand (1) the most important characteristics that are associated with software defects in the past; and (2) the most important characteristics that contributed to a prediction of a file.
Surprisingly, there exist numerous model-agnostic techniques to generate visual explanations (e.g., ANOVA and LIME) that have been used in the literature.
Particularly, we found that 18\% used ANOVA, 15\% used Variable Importance, 5\% used Partial Dependence Plot, 3\% used Decision Tree, 2\% used LIME, and 1\% used BreakDown to generate visual explanations.
Recently, Esteves~\ea~\cite{esteves2020mlsdp} also used SHAP~\cite{lundberg2017unified} to understand the predictions of defect prediction models.
Anchor~\cite{ribeiro2018anchors} (an extension of LIME~\cite{ribeiro2016should}) was proposed to present the visual explanations in the form of decision trees/rules.

Based on our analysis of the eight selected model-agnostic techniques (see Table~\ref{review-table}) that were used in prior studies, we found that visual explanations generated by these techniques produce different key information (e.g., important scores and relationship).
It remains unclear about which model-agnostic techniques are considered as the most preferred by practitioners to understand defect prediction models and their predictions.
Thus, we formulate the following research question:

\resultBox{(RQ2) \rqii}

\section{Survey Methodology}\label{section-design}


The goal of this work is to assess practitioners' perceptions of the goals of defect prediction models and the model-agnostic techniques for generating visual explanations of defect prediction models.
To address our two research questions, we conducted a qualitative survey study to investigate the practitioners' perceptions of the goals of defect prediction models and the model-agnostic techniques for generating visual explanations of defect prediction models.
We used a survey approach, rather than other qualitative approaches (e.g., interview), since we aim to assess their perceptions of the goals along 2 dimensions (i.e., perceived usefulness and willingness to adopt) and the model-agnostic techniques along 3 dimensions (i.e., overall preference, information usefulness, information insightfulness, and information quality).
Unlike an interview approach that is more unstructured, the closed-ended responses of the survey approach can be structured and quantified on a Likert scale which can be further analyzed to produce empirical evidence.
The open-ended responses of the survey approach also provide in-depth insights to synthesize and generate discussions.
As suggested by Kitchenham and Pfleeger~\cite{kitchenham2008personal}, we considered the following steps when conducting our study: (1) Survey Design (designing a survey and developing a survey instrument), (2) An Evaluation of the Survey Instrument (evaluating the survey instrument), (3) Participant Recruitment and Selection (obtaining valid data), (4) Data Verification (verifying the data), and (5) Statistical Analysis (analysing the data).
We describe each step below.

\subsection{Survey Design}

Our survey design is a cross-sectional study where participants provide their responses at one fixed point in time.
The survey consists of 9 closed-ended questions, 11 open-ended questions, and 1 one-ended question for feedback on our survey. 
The survey takes approximately 20 minutes to complete and is anonymous. 
Our survey can be found in the online supplementary materials~\cite{onlineappendix}.

To fulfil the objectives of our study, we created three sets of closed-ended and open-ended questions with respect to the demographic information, and the two research questions.
For closed-ended questions, we used agreement and evaluation ordinal scales.
To mitigate the inconsistency of the interpretation of numeric ordinal scales, we labeled each level of ordinal scales with words as suggested by Krosnick~\cite{krosnick1999survey}.
The format of our survey instrument is an online questionnaire.
We used Google Forms to implement this online survey.
When accessing the survey, each participant was provided with an explanatory statement which describes the purpose of the study, why the participant is chosen for this study, possible benefits and risks, and confidentiality.
Below, we present the rationale for the information that we captured:

\subsubsection*{Part 1--Demographics}

We captured the following information, i.e., Role: engineers, managers, and researchers; Experience in years (decimal value); Current country of residence; Primary programming language; Team Size: 1-10, 11-20, 21-50, 51-100, 100+; Usage of static analysis tools: Yes / No.


The collection of demographic information (i.e., roles, experience, country) about the respondents allows us to (1) filter respondents who may not understand our survey (i.e., respondents with less relevant job roles), (2) breakdown the results by groups (e.g., developers, managers, etc), and (3) understand the impact of the demographics on the results of our study.

Team size may have an impact on SQA practices.
For example, small teams might use a light-weight SQA practice (e.g., static analysis), while large teams might use a rigorous SQA practice (e.g., CI/CD and automated software testing).

Primary programming languages may impact SQA practices.
For example, some high-level programming languages might be easier to conduct SQA practices (e.g., Python and Ruby languages) than some low-level programming languages (e.g., C language).

The usage of static analysis tools may impact the practitioners' perceptions of the goals of defect prediction models and the model-agnostic techniques for generating visual explanations of defect prediction models.
For example, practitioners who use static analysis may not perceive the benefits of the prioritization goal of defect prediction models~\cite{wan2018perceptions}.
However, the ranking of the most risky files is not the only goal of defect prediction models.

\subsubsection*{Part 2--Practitioners' perceptions of the goals of defect prediction models}

To understand how practitioners perceive the goals of defect prediction models, we first illustrated the concept of defect prediction models then provided participants with the brief definition of each goal (as outlined in Section~\ref{sec:background}).
For each goal of defect prediction models, we assessed the practitioner's perceptions along two dimensions, i.e., perceived usefulness and willingness to adopt.
Perceived usefulness refers to the degree to which a person believes that using a particular system would enhance his or her job performance~\cite{lewis1992psychometric}
Willingness to adopt refers to the degree to which a person is willing to adopt a particular system~\cite{wan2018perceptions}. 
Thus, we asked the participants to rate the perceived usefulness and the willingness to adopt using the following evaluation ordinal scales: 

\begin{itemize}
    \item Perceived Usefulness: Not at all useful, Not useful, Neutral, Useful, and Extremely useful
    \item Willingness to adopt: Not at all considered, Not considered, Neutral, Considered, and Extremely considered
\end{itemize}

We then asked participants to describe the positive points and points for improvement about these goals of defect prediction models, and how the use of defect prediction models might impact their organizations when deploying in practice.

\subsubsection*{Part 3--Practitioners' perceptions of the model-agnostic techniques for generating visual explanations of defect prediction models}
We provided participants with examples of visual explanation that are generated from the 6 model-agnostic techniques for defect prediction models (i.e., VarImp, Partial Dependence Plots, Decision Tree, LIME, BreakDown, and Anchor).
We combined ANOVA and VarImp since both techniques provide the same information.
Similarly, we combined SHAP and BreakDown since both techniques provide the same information.
As suggested by Lewis~\ea~\cite{lewis1992psychometric,assila2016standardized}, we use the PSSUQ (Post-Study System Usability Questionnaire) framework to evaluate the practitioners' perceptions of the model-agnostic techniques for generating visual explanations of defect prediction models.
The PSSUQ framework focuses on four dimensions, i.e., information usefulness, information quality, information insightfulness, and the overall preference.
Information usefulness, information quality, and information insightfulness refer to the degree to which a person satisfies that using a particular visual explanation is useful, able to comprehend, and insightful to understand the characteristics that are associated with software defects and the characteristics that contributed to a prediction of a file, respectively.
For each dimension, we use the following evaluation ordinal scales: 
\begin{itemize}
    \item Extremely low, low, moderate, high, and extremely high. 
\end{itemize}

We then asked participants to describe the strengths and weaknesses of each visual explanation.
Finally, we asked an open-question to describe the ideal preferences of visual explanations for developing quality improvement plans.

To generate visual explanations for our survey, we used the release 2.9.0 and 3.0.0 of the Apache Lucene software system from Yatish~\ea~\cite{yathish2019affectedrelease}'s corpus.
The release 2.9.0 data (1,368 instances, 65 software metrics, and a defective ratio of 20\%) was used to construct defect prediction models, while the release 3.0.0 data (1,337 instances, 65 software metrics, and a defective ratio of 12\%) was used to evaluate such models to ensure that explanations are derived from accurate models.
We also used the release 3.0.0 data to generate visual explanations of LIME, SHAP, BreakDown, and Anchor.
To simplify the visual explanation for readability, we selected only five metrics, i.e., \texttt{AddedLOC}, \texttt{CommentToCodeRatio}, \texttt{LOC}, \texttt{nCommit}, and \texttt{nCoupledClass}.
We provided the steps for generating visual explanations in Zenodo~\cite{onlineappendix}.


\subsection{An Evaluation of the Survey Instrument}

We carefully evaluated our draft survey by using a pilot study for pre-testing~\cite{litwin1995measure}, prior to recruiting participants.
{We evaluated the survey with co-authors and PhD students who have background knowledge in software engineering research but may not restricted to the defect prediction domain.
They pointed out that the survey needs more context and details, especially for non-domain experts. 
Particularly, the draft survey did not provide the definition of software defect prediction, how they are handled in software companies, and how defect prediction models are used to support decision-making.
Thus, in the beginning of Sections 3 and 4 of the revised draft survey, we included overview figures and scenario-based explanations to address the concern.
We repeatedly refined the survey instrument to identify and fix potential problems (e.g., missing, unnecessary, or ambiguous questions) until reaching the consensus among the pre-testing participants.}
Finally, the survey has been rigorously reviewed and approved by the Human Research Ethics Committee of our university.

\subsection{Participant Recruitment and Selection}

The target population of our study is software practitioners.
To reach our target population, we used the recruiting service as provided by the Amazon Mechanical Turk.
Unlike StackOverflow or Linkedin, the Amazon Mechanical Turk platform comes with built-in options to filter participants for participant selection and customize a monetary incentive for each participant.
Particularly, we applied the participant filter options of ``\emph{Employment Industry - Software \& IT Services}'' and ``\emph{Job Function - Information Technology}'' to ensure that we reached the target population.
We paid 6.4 USD as a monetary incentive for each participant~\cite{smith2013improving,edwards2002increasing}.
In total, our survey has 9 closed questions (450 responses) + 11 open questions (550 responses) + 1 open question (50 responses) for feedback.

\subsection{Data Verification}

We manually read all of the open-question responses to check the completeness of the responses i.e., whether all questions were appropriately answered.
We excluded 68 responses that are missing and are not related to the questions.
In the end, we had a set of 982 responses.
We summarized and presented the results of closed-ended responses in a Likert scale with stacked bar plots, while we discussed and provided examples of open-ended responses.



\begin{figure*}[t]
\centering
\begin{subfigure}[t]{.48\hsize}
\includegraphics[width=\columnwidth]{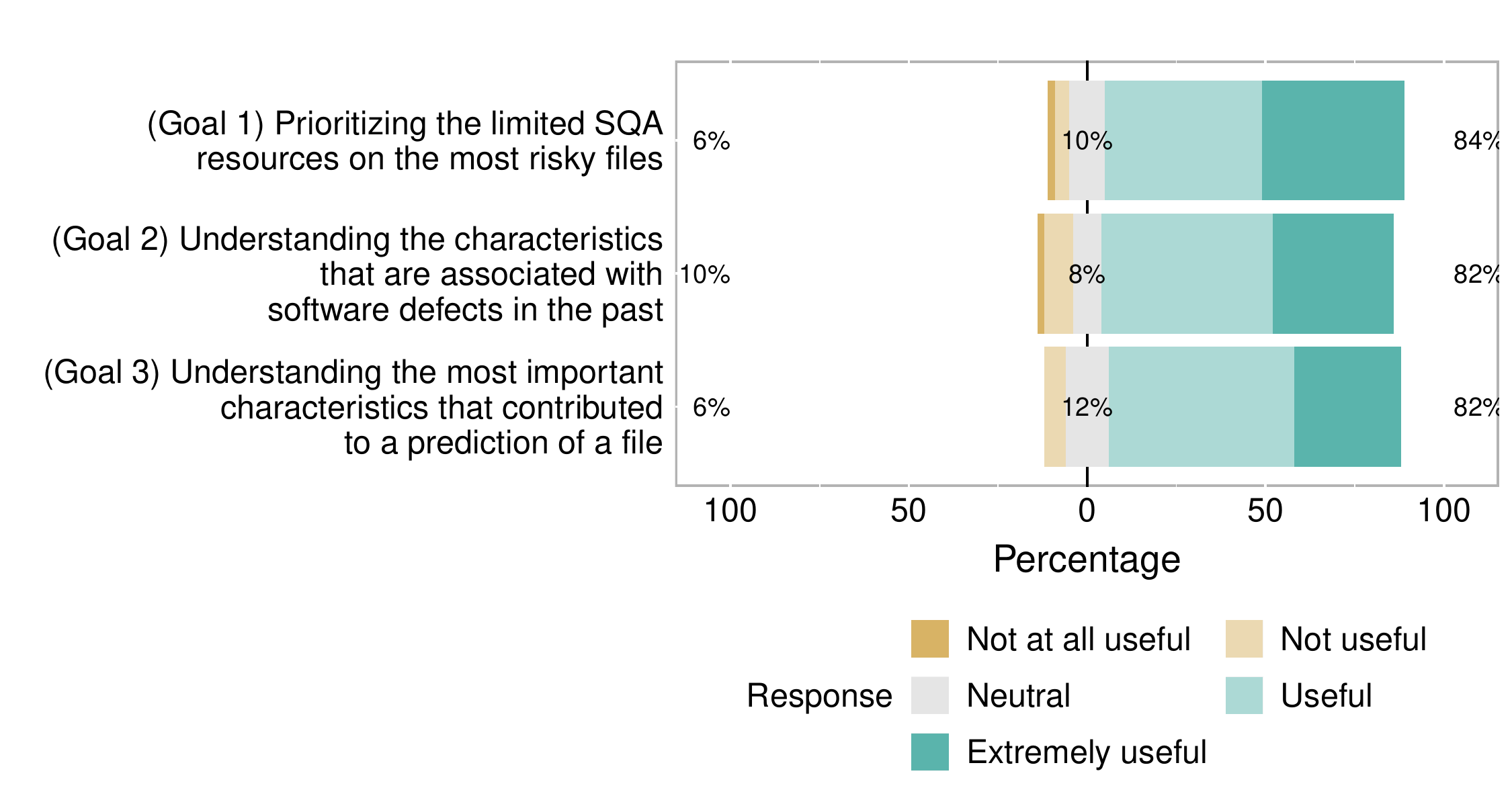}
\caption{Perceived Usefulness}
\label{rq2-figure-1-1}
\end{subfigure} 
\hfill
\begin{subfigure}[t]{.48\hsize}
\includegraphics[width=\columnwidth]{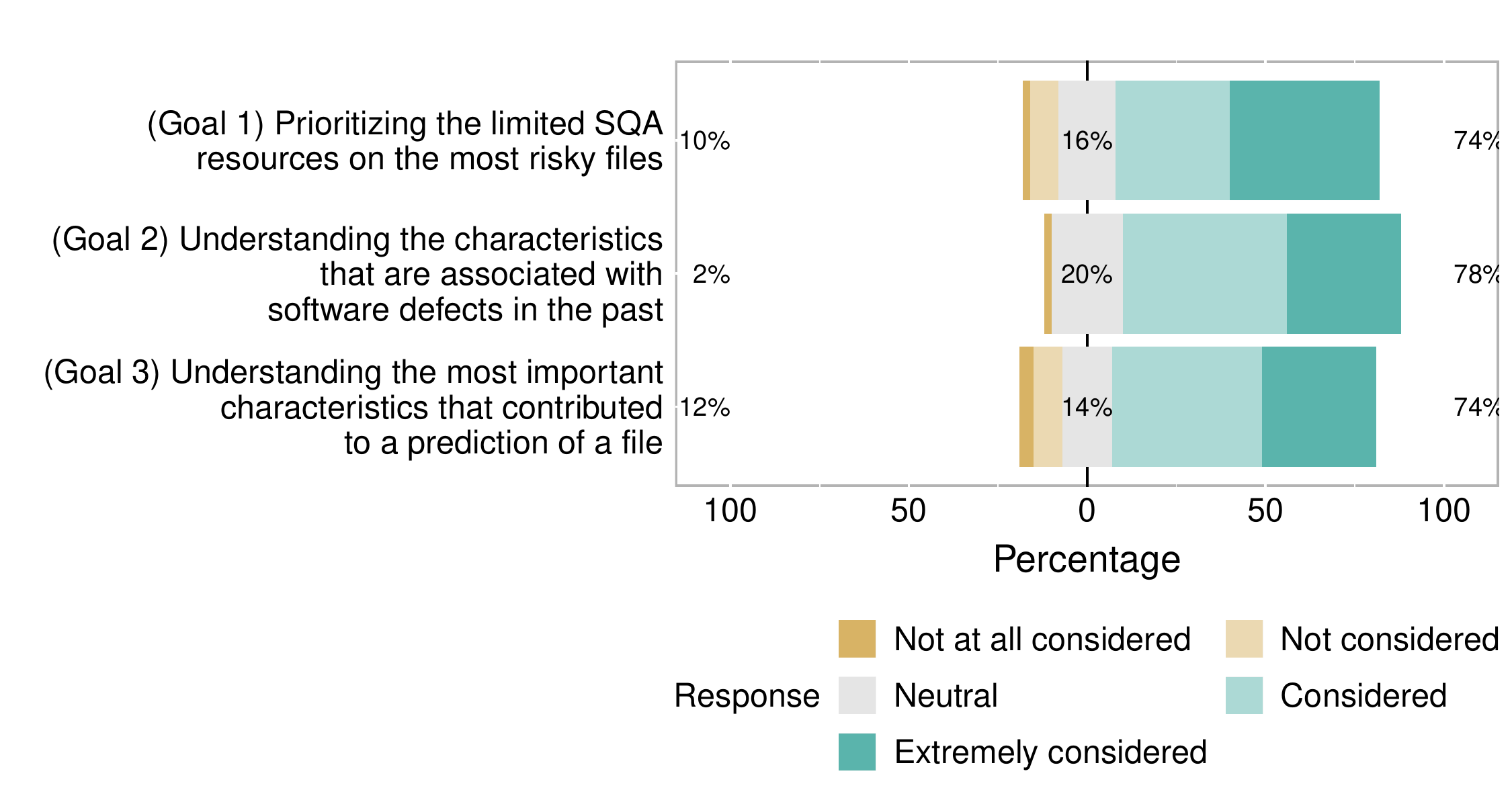}
\caption{Willingness to adopt}
\label{rq2-figure-1-2}
\end{subfigure} 
\caption{The likert scores of the perceived usefulness and the willingness to adopt from the respondents for each goal of defect prediction models.}
\label{fig:rq1-results}
\end{figure*}

\subsection{Statistical Analysis}

For the closed-end questions with ordinal scales, we converted the ratings into scores.
For example, we converted not at all useful, not useful, neutral, useful, and extremely useful to 1, 2, 3, 4 and 5 respectively. 
Then, we applied the ScottKnott ESD test to clusters of distributions into statistically distinct ranks.
We used the implementation of the ScottKnott ESD test as provided by the ScottKnottESD R package~\cite{tantithamthavorn2016scottknottesdRpackge,tantithamthavorn2017empirical,tantithamthavorn2018optimization}.

For ratings of statements, we calculated the percentage of respondents who strongly agree or agree with each statement (\% strongly agree+\% agree) and the percentage of respondents who strongly disagree or disagree with each statement (\% strongly disagree+\% disagree). 
As suggested by Wan~\ea~\cite{wan2018perceptions}, we also computed an agreement factor for each statement. 
The agreement factor is a measure of agreement between respondents, which is calculated for each statement by the following equation: (\% strongly agree + \% agree)/(\% strongly disagree + \% disagree).
High values of agreement factors indicate a high agreement of respondents to a statement.
The agreement factor of 1 indicates that the numbers of respondents who agree and disagree with a statement are equal.
Finally, low values of agreement factors indicate that a high disagreement of respondents to a statement.

\section{Survey Results}\label{section-results}

We present the demographics of our survey, and then the results of using survey data to answer our research questions.

\subsection{\textbf{Demographics}}
The top two countries in which the respondents reside are India (58\%) and the United States (36\%).
Among the respondents, they described their job roles as:
Developers (50\%), Managers (42\%), and others (8\%).
The number of years of professional experience of the respondents varied from less than 5 years (26\%), 6--10 years (38\%), 11--15 years (22\%), 16--20 years (12\%), and more than 25 years (2\%).
They described their team size as: less than 10 people (30\%), 11--20 people (30\%), 21--50 people (26\%), 51--100 people (2\%), and more than 100 people (12\%).
The respondents described their experience in programming languages as: Java (44\%), Python (30\%), C/C++/C\# (28\%), and JavaScript (12\%).
They also answered whether they are using static analysis tools in their organizations as follows: Yes (62\%) and No (38\%).

These demographics indicate that the responses are collected from practitioners resided in various countries, roles, years of experience, and programming languages, indicating that our findings are likely not bound to specific characteristics of practitioners.


\begin{table*}[t]
\centering
\footnotesize
 \caption{(RQ1) A summary of the ScottKnott ESD rank, the agreement percentage, the disagreement percentage, and the agreement factor for the three goals of defect prediction models.}
\label{tab:rq1-table}
\resizebox{.8\textwidth}{!}{
\begin{tabular}{c||l|c|c|c|c}

\textbf{Dimension}                                                                                & \textbf{Goal} & \textbf{SK Rank} & \textbf{\% Agreement} & \textbf{\% Disagreement} & \textbf{Agreement Factor}                \\ \hline\hline
\multirow{7}{*}{\textbf{\begin{tabular}[c]{@{}c@{}}Perceived\\ Usefulness\end{tabular}}} 
& {{\begin{tabular}[c]{@{}l@{}}Goal 1 -- Prioritizing the limited SQA\\resources on the most risky files\end{tabular}}}    & 1                   & 84\%   & 6\%                       & 14.00                     \\ \cline{2-6}
& {{\begin{tabular}[c]{@{}l@{}}Goal 2 -- Understanding the characteristics\\that are associated with software defects\\   in the past\end{tabular}}} & 1                  & 82\%   & 10\%                       & 8.20                                                 \\ \cline{2-6}
& {{\begin{tabular}[c]{@{}l@{}}Goal 3 --  Understanding the most\\important characteristics that contributed\\to a prediction of a file\end{tabular}}}      & 1    & 82\%     & 6\%                     & 13.67                                                     \\ \hline\hline
\multirow{7}{*}{\textbf{\begin{tabular}[c]{@{}c@{}}Willingness\\ to Adopt\end{tabular}}}                                          
& {{\begin{tabular}[c]{@{}l@{}}Goal 1 -- Prioritizing the limited SQA\\resources on the most risky files\end{tabular}}}       & 1      & 74\% & 10\% & 7.40                      \\ \cline{2-6}
& {{\begin{tabular}[c]{@{}l@{}}Goal 2 -- Understanding the characteristics\\that are associated with software defects\\   in the past\end{tabular}}}  & 1                      & 78\%    & 2\%                      & 39.00                      \\ \cline{2-6}
& {{\begin{tabular}[c]{@{}l@{}}Goal 3 --  Understanding the most\\important characteristics that contributed\\to a prediction of a file\end{tabular}}}        & 1           & 74\%    & 12\% & 6.17                                                      \\ \cline{2-6}
\hline
\end{tabular}}
\end{table*}

\subsection{\textbf{\rqi~(RQ1)}} 
Figure~\ref{fig:rq1-results} presents the Likert scores of the perceived usefulness and the willingness to adopt from the respondents for each goal of defect prediction models.
Table~\ref{tab:rq1-table} presents a summary of the ScottKnott ESD rank, the agreement percentage, the disagreement percentage, and the agreement factor for the three goals of defect prediction models. 

\textbf{82\%-84\% of the respondents perceived that the three goals of defect prediction models are useful and nearly 80\% of them are willing to adopt.}
Figure~\ref{fig:rq1-results} shows that 82\%-84\% and 72\%-78\% of respondents rate that the goals of defect prediction models are perceived as useful and considered willing to adopt, respectively.
Table~\ref{tab:rq1-table} also confirms that the agreement factors are high across all goals with the values of 8.2-14 and 6-39 for perceived usefulness and willingness to adopt, respectively.
The high agreement factors of responses provided by the respondents suggest that most respondents provide positive responses (e.g., useful and extremely useful) when comparing negative responses (e.g., not useful and not at all useful).
Respondents provided rationales that if defect prediction models were adopted, they are likely to save developers' effort, e.g., (R6: ``\textit{... saves developers a huge amount of effort on reviewing or testing non-defective files ...}''), and improve the efficiency of code inspection (R8: ``\textit{Issues can be caught early in development.}'', (R24: ``\textit{More time will be focused on critical areas. 
Less time will be wasted on areas without defects.}'').
The ScottKnott ESD test also ranks all of the goals at the same rank, confirming that the scores among the goals have negligible effect size difference.
This finding highlights that \emph{not only the defect prediction goal but also the other two goals (i.e., understanding defect prediction models and their predictions) receive similar perceptions of usefulness and willingness to adopt with no statistically significant difference}.


Below, we discuss further if the respondents' demographics have any impact on their perceptions.


\textbf{The use of static analysis tools has no significant impact (with a negligible to small effect size) on their willingness to adopt defect prediction models that are developed from various goals.}
The emergence of static analysis and defect prediction models is in parallel with different intellectual thoughts: one is driven by algorithms and abstraction over code, while defect prediction models are driven by statistical methods over large defect datasets~\cite{wan2018perceptions}. 
Wan~\ea~\cite{wan2018perceptions} noted that static analysis shares some overlapping goals with defect prediction models, i.e., improving inspection efficiency, finding minimal, and potentially defective regions in source code.
Thus, practitioners who use static analysis may not be willing to adopt defect prediction models.
In contrast, we did not observe any significant impact of the use of static analysis tools on their willingness to adopt defect prediction models.
This finding is aligned with Rahman~\ea~\cite{rahman2014comparing} who found that both static analysis and statistical defect prediction models provide comparable benefits.


We found that \textbf{Team Size has the largest influence on their willingness to adopt.}
To investigate the impact of various demographic factors on their willingness to adopt, we built a linear regression model by using the \texttt{ols} function of the \texttt{rms} R package. 
The independent variables are the years of experience, roles, team size, programming languages, and static analysis, while the dependent variable is the willingness score.
After using the optimism-reduced bootstrap validation (i.e., a model validation technique that randomly draws training samples with replacement then tests such models with original samples and the samples used to construct these models), the regression model achieves a goodness-of-fit ($R^2$) of 0.35.
Then, we analyzed the Chi-square statistics of the ANOVA Type-II analysis, then normalized these Chi-square statistics into percentage to better illustrate the relative differences among variables.
The ANOVA analysis indicates that Team Size has the largest influence on their willingness to adopt (i.e., 52.80\% for TeamSize, 20.98\% for useJava, 13.05\% for usePython, 7.92\% for Year, 5.01\% for role).
We found that the respondents who use Java, with little years of experience and a large team size (more than 100 people) tend to consider willing to adopt defect prediction models.
Nevertheless, we observe a minimal impact of the roles (developers vs managers) on their perceptions.
We provided a detailed analysis of marginal effect size of each demographic factor on the estimated willingness to adopt defect prediction models in Zenodo~\cite{onlineappendix}.
\begin{table*}[t]
\centering
\caption{(RQ2) A summary of the ScottKnott ESD rank, the agreement percentage, the disagreement percentage, and the agreement factor for each model-agnostic technique for generating visual explanations of defect prediction models.}
\label{table_rq2_viz_results}

\begin{footnotesize}
\begin{tabular}{c||l|c|c|c|c}

\textbf{Dimension}                       & \textbf{Techniques}  & \textbf{SK Rank} & \textbf{\%Agreement} & \textbf{\%Disagreement} & \textbf{Agreement Factor}      \\ \hline\hline
\multirow{6}{*}{\textbf{Usefulness}}     & \textbf{LIME}     & \textbf{1}       & \textbf{76\%}  & \textbf{6\%}               & \textbf{12.67}          \\ \cline{2-6}
& ANOVA/VarImp         & 2                & 60\%   & 14\%                         & 4.29                                                       \\ \cline{2-6}
& PDP              & 2                & 60\%    & 18\%                        & 3.33                                                       \\ \cline{2-6}
& BreakDown/SHAP         & 2                & 50\%   & 18\%                         & 2.78                                                       \\ \cline{2-6}
& Decision Tree         & 2                & 54\%   & 18\%                         & 3.00                                                       \\ \cline{2-6}
& Anchor/LORE         & 2                & 60\% & 28\%                            & 2.14                                                       \\ \hline\hline
\multirow{6}{*}{\textbf{Insightfulness}} & \textbf{LIME}     & \textbf{1}       & \textbf{68\%}  & \textbf{8\%}                & \textbf{8.50}             \\ \cline{2-6}
& \textbf{Decision Tree}  & \textbf{1}       & \textbf{52\%} & \textbf{10\%}                 & \textbf{5.20}                                              \\ \cline{2-6}
& BreakDown/SHAP         & 2                & 58\%    & 12\%                        & 4.83                                                       \\ \cline{2-6}
& PDP                 & 2                & 54\%  & 16\%                          & 3.38                                                       \\ \cline{2-6}
& ANOVA/VarImp         & 2                & 58\%     & 18\%                       & 3.22                                                       \\ \cline{2-6}
& Anchor/LORE        & 3                & 46\%   & 24\%                         & 1.92                                                       \\ \hline\hline
\multirow{6}{*}{\textbf{Quality}}  & \textbf{LIME}      & \textbf{1}       & \textbf{66\%}  & \textbf{4\%}                 & \textbf{16.50}            \\ \cline{2-6}
& \textbf{ANOVA/VarImp}     & \textbf{1}       & \textbf{70\%} & \textbf{8\%}                   & \textbf{8.75}                                          \\ \cline{2-6}
& \textbf{Decision Tree}    & \textbf{1}                & \textbf{70\%}  & \textbf{14\%}                          & \textbf{5.00}                                                       \\ \cline{2-6}
& PDP       & 2                & 56\%       & 24\%                     & 2.33                                                       \\ \cline{2-6}
& BreakDown/SHAP        & 2                & 52\%  & 26\%                          & 2.00                                                       \\ \cline{2-6}
& Anchor/LORE      & 2                & 56\%         & 24\%                   & 2.33                                                    \\ \hline\hline
\end{tabular}
\end{footnotesize}
\end{table*}

\subsection{\textbf{ \rqii~(RQ2)}}\label{section-rq3}

\textbf{LIME is the most preferred model-agnostic technique to understand the most important characteristics that contributed to a prediction of a file with an agreement percentage of 66\%-78\% along three dimensions.} 
As shown in Table~\ref{table_rq2_viz_results}, LIME consistently appears at the top-1 ScottKnott ESD rank for all three dimensions with an agreement percentage of 76\% for information usefulness, an agreement percentage of 68\% for information insightfulness, and an agreement percentage of 66\% for information quality.
Respondents found that LIME is very easy to understand, e.g., R1 (\emph{``Risk scores are visually oriented and users will understand it faster.''}) and R50 (\emph{``Easy to understand, we just want to know what's already ok and what's need to improve.''}). 
However, some respondents raised concerns that LIME generates too much information (i.e., too many characteristics for many defective files), e.g., R9 (\emph{``... But it also could be time consuming to review depending on if there are many different files to review.''}) and R50 (\emph{``... too large and too tired to read 10 charts for 10 files.''}).

\textbf{ANOVA/VarImp is the second most preferred technique to understand the characteristics that are associated with software defects in the past with an agreement percentage of 58\%-70\% along three dimensions.}
As shown in Table~\ref{table_rq2_viz_results}, ANOVA/VarImp consistently appears at the second ScottKnott ESD Rank, except for information quality, with an agreement percentage of 60\% for information usefulness, an agreement percentage of 58\% for information insightfulness, and an agreement percentage of 70\% for information quality.
Similar to LIME, respondents found that the bar charts of ANOVA/VarImp are very easy to understand, e.g., R50 (\emph{``Explains risk values of several factors.''}).
This finding indicates that while both LIME and ANOVA/VarImp generate different information, they are complementary to each other.
This suggests that while LIME should be used to understand the most important characteristics that contributed to a prediction of a particular file (Goal 3), ANOVA/VarImp is still needed to understand the overview of the general characteristics that are associated with software defects in the past (Goal 2).


\section{Threats to  Validity}\label{section-threats}

\textbf{Construct validity:}
We studied a limited period of publications (i.e., 2015-2020).
Thus, the results may be altered if the studied period is changed.
Future research should consider expanding our study to a longer publication period.

Hilderbrand~\ea~\cite{hilderbrand2020engineering} found that there are statistically gender differences in the cognitive style of developers.
Yet, gender is not considered in our survey.
Thus, our recommendation may not be generalized to all genders.
Future studies should consider gender aspects when collecting the demographics of respondents.

In this paper, we design the survey with the assumptions of file-level defect prediction. 
However, Wan~\ea~\cite{wan2018perceptions} found that commit level is the most preferred by practitioners.
Thus, our results may not be applicable to other granularity levels of predictions (e.g., commits, methods).




\textbf{Internal validity:} One potential threat is related to the bias in the responses due to the imbalanced nature of the recruited participants.
Also, practitioners' perceptions may biased and can change from one person to another or even one organization to another~\cite{devanbu2016belief}.
However, the population of the recruited participants is composed of practitioners of different roles, years of experience, country of residence, and programming languages.
To mitigate issues of fatigue bias in our survey study, we conducted a pilot study with co-authors and PhD students to ensure that the survey can be completed within 20 minutes.

\textbf{External validity:} We recruited a limited number of participants.
Thus, the results and findings may not generalise to all practitioners.
Nevertheless, we described the survey design in details and provided sets of survey questions in the online supplementary materials~\cite{onlineappendix} for future replication.
\section{Lessons Learned and Open Questions}

We discuss key lessons learned from the results of our qualitative survey and discuss some open questions for developing the next-generation of defect prediction models.
We hope that such open questions can motivate important future work in the area and foster the adoption of the defect prediction models in practice.

\textbf{(Open Question-1) How can we improve the efficiency and effectiveness of SQA planning?}
Over the past decade, there have been hundreds of studies focusing on improving the prediction goal of defect prediction models~\cite{hall2012systematic}.
However, as we discussed in Section~\ref{sec:background}, prediction is not the sole goal of defect prediction models.
Instead, defect prediction models can also help practitioners understand key important characteristics that are associated with software defects in the past to develop SQA plans.
As discussed in answering RQ1, practitioners also perceived that understanding defect prediction models and their predictions are equally useful with no statistically significant difference.
For example, R28 highlights the benefits of data-informed knowledge sharing as follows: \textit{``Management can provide suggestions (derived from defect prediction models) to improve the software quality (to developers) and to improve the process and policy.''}
Thus, future studies should start exploring a new research direction on \emph{``improving the efficiency and effectiveness of SQA planning,''} (e.g., \cite{rajapaksha2021sqaplanner}) in addition to saving code inspection effort.

\textbf{(Open Question-2) What is the most effective visual explanation for developing SQA plans that address practitioners' needs?}
The results of answering RQ1 (see Table~\ref{tab:rq1-table}) confirm that our respondents perceived that understanding defect prediction models and their predictions (i.e., Goal 2 and Goal 3) are both useful and considered willing to adopt them. 
However, the results of answering RQ2 (see Table~\ref{table_rq2_viz_results}) show that 6\%-28\% of the respondents do not perceive that visual explanations generated by model-agnostic techniques as useful, insightful and comprehensive.
In particular, while we find that some respondents rated LIME as the most preferred model-agnostic technique (i.e., Rank 1$^{st}$) to understand the most important characteristics that contributed to a prediction of a file, some respondents rated LIME as the least preferred visual explanation (i.e., Rank 6$^{th}$) (e.g., R6: \textit{``Too confusing to me.''}).
This finding suggests that none of the model-agnostic techniques can satisfy all software practitioners and their needs.
This is because model-agnostic techniques are originated from the explainable AI domain and are not designed for software engineering.
Thus, future studies should start exploring a new research direction on \emph{``inventing a domain-specific and human-centric visual explanation for helping practitioners to develop SQA plans.''}

\textbf{(Open Question-3) How can we improve the scalability of defect prediction models at the lowest implementation cost, while maintaining their explainability?}
Defect prediction models have been proposed over several decades. 
However, the deployment of such models is still limited to top software organizations, e.g., Microsoft, Google, and Cisco.
Respondents raised concerns about the cost of the implementation and deployment of defect prediction models (e.g., R28: ``\textit{... However, I think the implementation of defect prediction models can be a challenging task and it is likely to be time-consuming and expensive.}'').
It is widely known that the implementation of defect prediction models requires a deep understanding of the best practices for mining, analyzing and modelling software defects~\cite{tantithamthavorn2018pitfalls}.
Thus, future studies should focus on applying this best practice in a larger-scale experiment, i.e. shifting the research direction from analytics-in-the-small to analytics-in-the-large.
Based on our experience, few challenges are very important to address when increasing the scalability of defect prediction models, e.g., how to ensure that defect prediction models are compatible across development environments e.g., programming languages and operating systems.
This is similar to one respondent who stated that (R34: ``\textit{Ensure that it is language independent and works on multiple operating systems. Legacy systems will obviously pose a challenge.}'').

In addition, when developing defect prediction models in-the-large (i.e., learning from multiple projects), the same set of metrics are still required.
However, in reality, different software projects use different software development tools, e.g., Jira, Git, Gerrit, Travis.
Prior studies had attempted to build a universal defect model~\cite{zhang2014towards} and deal with heterogeneous data~\cite{nam2017heterogeneous}.
However, there are other challenges when dealing with heterogeneous data across software development tools and ultra-large-scale modelling approaches.
Future studies should explore the development of \emph{a highly-accurate and scalable (pre-trained) defect analytics model at the lowest implementation cost while maintaining explainability}.

\newpage 

\textbf{(Open Question-4) How can we assure developers' privacy and fairness when deploying defect prediction models in practice?}
One of the respondents (R3) raised important concerns about a lack of \emph{``privacy''} when using defect prediction models.
For example, Facebook currently has an Ownesty system that aims to identify who is the most suitable owner of a given asset changes over time, i.e. code ownership~\cite{ahlgren2020ownership}.
Prior studies show that code ownership metrics share a strong relationship with defect-proneness~\cite{Bird2011a}.
Thus, many practitioners are afraid of a lack of privacy and fairness of defect prediction models.
In particular, would developers be laid-off due to the use of defect prediction models for identifying who introduce software defects?
Thus, managers should carefully handle expectations with developers about the usage policy and the code of conduct when adopting defect prediction models in practice.

\textbf{(Open Question-5) How can we best enable human-in-the-loop when using defect prediction models?}
Researchers have proposed lots of automated tools in software engineering in order to greatly improve the developers' productivity and software quality.
However, one respondent mentioned that such automated tools may have a negative impact on the development behaviors of developers (R6: ``\textit{Laziness and do not care with catching errors and dumping all the work on explainable defect (prediction) models.}'').
Importantly, this concern indicates that the adoption of automated tools may lead to a lax software development process e.g. solely relying on the tools rather than developers' skills.
Thus, it is important to manage practitioners' expectations that such automated tools should be used as guidance to support decision-making and policy-making, not replacing developers' jobs.
Therefore, the next-generation defect prediction models should focus on how to enable human-in-the-loop into the defect prediction models, combining the best of human intelligence with the best of machine intelligence.


\vspace{-2mm}

\section{Conclusions}\label{section-conclusion}

In this paper, we presented the findings of the qualitative survey of 50 practitioners on their perceptions of the goals of defect prediction models and the model-agnostic techniques for generating visual explanations for defect prediction models. 
We conclude that:
(1) Researchers should put more effort on investigating how to improve the understanding of defect prediction models and their predictions, since our analysis of related work found that these two goals are still under research despite receiving similar perceptions of usefulness and willingness to adopt with no statistically significant difference; and 
(2) Practitioners can use LIME and ANOVA/VarImp to better understand defect prediction models and their predictions.
Finally, we discuss many open questions that are significant, yet remain large unexplored (e.g., developers' privacy and fairness when deploying defect prediction models in practice, and human-in-the-loop defect prediction models).



\textbf{Acknowledgement.} CT was supported by ARC DECRA Fellowship (DE200100941).
JG was supported by ARC Laureate Fellowship (FL190100035).

\bibliographystyle{IEEEtranS}
\bibliography{filteredref.bib} 

\end{document}